\documentstyle[prd,tighten,aps]{revtex}
\begin{document}
\draft

\twocolumn[\hsize\textwidth\columnwidth\hsize\csname
@twocolumnfalse\endcsname
\renewcommand{\theequation}{\thesection . \arabic{equation} }
\title{\bf Representations of relativity, quantum 
gravity and cosmology}

\author{Pedro F. Gonz\'alez-D\'{\i}az}
\address{Centro de F\'{\i}sica ``Miguel Catal\'an'',
Instituto de Matem\'aticas y F\'{\i}sica Fundamental,\\
Consejo Superior de Investigaciones Cient\'{\i}ficas,
Serrano 121, 28006 Madrid (SPAIN)}
\date{December 20, 1997}

\maketitle

\begin{abstract}

We review an attempt to set a suitable foundational
principle for
consistent quantization of gravity, based on the canonical
formulation. It requires extending the spacetime description
of the relativistic postulates to also encompass an alternative
formulation in momentum-energy continuum where the inertial
physical laws can be equivalently described. The extension to
noninertial frames breaks such an equivalence, leaving a new
dynamical field which, together with gravity, allows to
construct a canonical scenario where the Dirac's quantization
method leads to consistent definitions of hermitian ordering
for the operators of the canonical quantum theory.

\end{abstract}


\vskip2pc]

\renewcommand{\theequation}{\arabic{section}.\arabic{equation}}

\section{\bf Introduction}

It is a well-known fact that relativity relegates space and time
to the subjective role of the elements of a language that any
observer may use to describe the laws that govern the behaviour
of the objective reality when this is interpreted as a set of
matter bits. Actually, it is the relation between the only two
kinds of the objetivized entities allowed by the theory -namely
the observer and the bits of matter, that one must take as the
fundamental building blocks of relativity, and it makes no
difference to the predictions of the theory which part of the
objective physical system is identified with the observer and
which part is called an observed bit of matter.

Objectivizing bits of matter is nevertheless just a particular
way to analyse physical reality along the task of dividing it
into its minutest pieces. There still exists another
approach to look at the physical reality that manner. It
consists of objectivizing bits of space and time rather than
matter, and taking them as the fundamental constituent parts
of reality. Adopting such an approach would be allowed from
two fundamental developments in gravitational physics. One
of them is the realization that there exist very small, but
still nonzero fundamental length and time, possibly at
the Planck scale, which determine the finite maximum
resolution for all experiments. The other development is
relativistic cosmology itself. Here, physical reality can
be looked at as being described by the set of relations 
between distances and times that characterize the large
scale structure of the universe. In its canonical formulation,
moreover, the dynamical content of cosmology is given by
the Hamiltonian constraint which, being zero, is prepared
to be treated both as an energy-momentum object, or as
a space-time object, depending on whether we divide or
multiply by the Planck-length squared. In the latter case,
it would be momentum and energy
which should be relegated to the sujective role of the elements
of a mere language that the observers would use to describe
a physical reality made up of objectivized bits of space and
time, with the building blocks of the theory now being the
relation between observers and such objectivized bits. One
would expect that it again makes no difference to the
predictions of the theory in this representation which
part of the physical system is identified with the observer
and which part is taken to play the role of an observed bit
of space or time.

Clearly, the kinematics of Einstein relativity for inertial
systems gives, through Lorentz transformations, the relativistic
changes of time durations and space distances that an observer
may measure with clocks and meters. It appears then that for
momentum and energy to be relegated to elements of a subjective
language, they should enter a relativistic formalism formally
identical to that for spacetime relativity and reproduce the
same transformations for momentum and energy, and space and
time as well, though the latter two quantities of such a
formalism should be taken as objective elements of the
physical reality.

On the other hand, a traditional debate about wave mechanics
refers to whether it requires relativity theory to be
consistently formulated [1]. Of course, de 
Broglie derived his known
relation based on a relativistic foundation [2]. In order to 
explain x-ray diffraction in crystals by means of the
corpuscular theory of light though, Duane postulated [3] the
momentum rule before de Broglie without relativistic
foundation. But, as formulated by Duane, the rule becomes
obscure without de Broglie's idea of the correspondence of
particle and wave, and without his relativistic proof that
the group velocity of a wave packet coincides with the
velocity of the corresponding particle [4]. Thus, if we adhere
to the currently most accepted view that wave mechanics
is only {\it derivable} from relativistic concepts
and has an in principle well-defined nonrelativistic limit, then
one could formulate a wave mechanics which would follow
from the alternate relativistic approach relegating momentum
and energy to the role of subjective quantities. Since
the above two relativistic representations 
should be expected to be equivalent
while keeping within the inertial framework,
one would also expect the two resulting formulations of
wave mechanics to be equivalent for inertial 
relativistic frames.

The idea that we shall explore in this paper is: when we
generalize to noninertial frames, besides the same energy,
momentum, time and space intervals, these relativistic
representations should give rise to two generally {\it distinct}
inequivalent dynamical field quantities -namely the usual
gravitational field and a new field which we interpret as
describing cosmological interactions without nonrelativistic
counterpart. Clearly, Einstein's general relativity 
has proved successful in dealing with the cosmological
problem under the special restrictions about spatial
homogeneity and isotropy of matter in the universe implied
by the cosmological and Weyl principles. Our claim is that
the spacetime relativistic description of cosmology is only
valid in the approximation of large universes, but as one goes
back to the earliest stages of the cosmological evolution it
is the cosmological field which dominates the dynamics of
the evolution. In turn, when the universe becomes large enough,
the cosmological interactions considered here 
are by themselves alone not adequate to describe
the universe. Actually, what would exactly describe the
evolution of the universe at any stage is the combined effect
of the Einstein equations and the equations for the
introduced cosmological field.
A consequence from this point of view is that neither the
gravitational nor the cosmological field can be quantized
separately in a fully consistent way, being the above combined
effect of the two fields which admits a full quantization free
from the usual problems of the canonical formalism of quantum
gravity.

In Sec. II we consider in detail the momentum-energy formulation
of spacial relativity. It is shown how this and Einstein
relativity can be both derived from a generalized abstract
relativistic formalism where action coordinates are used.
The wave-mechanical implications from such a relativistic
approach is also dealt with in this section. An extended
canonical formalism for noninertial frames is discussed in
Sec. III, where we see how the gravitational and cosmological
fields can be combined in a unique picture. Sec. IV deals
with the quantization of the generalized canonical formalism.
A proof is given that the resulting quantum approach does not
show any problem with the hermitian order of operators.

\section{\bf Representations of relativity}
\setcounter{equation}{0}

There is one sense in which quantum-mechanical 
position and momentum representations are not 
formally equivalent if wave-particle
duality is, as usual, meant to imply equal contributions from
the two pictures (wave and particle) to that duality. When one 
presupposes a system to be an elementary particle with mass $m$,
the particle is being assumed to be point-like and its wave
function in $p$-representation $\Psi(p)$ can also be written
as a function of the wavelength, namely $\Psi(\lambda)$, by
using the de Broglie relation $p=\frac{h}{\lambda}$, which also
holds in the nonrelativistic limit.
Therefore, $\Psi(p)=\Psi(\lambda)$ can always be interpreted as the
probability amplitude for the presupposed
particle-like system to behave as a wave with
wavelength $\lambda$. At the same time, in $x$-representation,
$\Psi(x)$ is regarded as the probability amplitude for the
presupposed point-like particle to be localized in space at $x$. However,
if one would presuppose the system to be a wave with wavelength
$\lambda$, whereas $\Psi(p)$ could equivalently be regarded as
the probability amplitude for the wave to propagate with a
momentum ``localized'' at the single value $p$, 
there is no known fundamental quantum relation
allowing the spatial distance $x=R$ in
$\Psi(x)$ (with $R$ being the objectivized bit of spatial
distance characterizing the system in wave representation)
to be discretized so that this wave function can be
re-written as a function of a
corresponding particle property
(which we take to be the mass), namely $\Psi(m)$, 
interpretable as the probability amplitude for the wave-like 
system to
behave like a particle with mass $m$.

In the relativistic formalism, $x$ could still be discretized in
terms of a relativistic Compton wavelength of the system,
$R\equiv\lambda_c=\frac{h}{mc}$, i.e. in terms of the
spatial scale at which the system undergoes purely relativistic
interactions with effects
such as the fine-structure originating
from its spin. However, this relation would be lost in the
limit $c\rightarrow\infty$, where $R\rightarrow 0$, and in any
case, cannot be considered as a {\it fundamental} quantum
relation that could be regarded to be at the same footing as
the de Broglie formula, in this case relating a measurable bit of
objectivized space distance to mass. Moreover, even at the
relativistic level, there exists no known quantum relation
whatsoever which would link a discretized bit of
objectivized time, $T$ (characterizing the system), to the
mass of that system, leading to a transformation $\Psi(T)
\rightarrow\Psi(m)$, analogous to as the Einstein-de Broglie
relation $E=h\nu$ does with energy and frequency to allow the
transformation $\Psi(E)\rightarrow\Psi(\nu)$.

Although, given the mass of the electron, it is our choice
whether to measure its position or momentum, and this is
still enough to describe objective reality in the inertial
approximation, the alluded inequivalence appears to be
detrimental to the beauty of the underlying theory and
leads, in fact, to the known difficulties encountered in
any attempt to quantize gravity (see Subsec. II-D and Secs.
III and IV). The electron has a mass, but e.g. in experiments
where its interaction with the Coulomb field of the hydrogen
nucleus is measured, it also shows another element of its
objective reality which, like mass, only depends on relative
velocity (through the relativistic factor): the spatial
domain given by the Compton wavelength where the Darwin
interaction takes place, or equivalently, the time interval
that a train a light waves would take in traversing that
spatial domain.

The lack of a fundamental quantum relation between $R$ and $m$
and between $T$ and $m$ leading 
to the above formal inequivalence 
appears to be related to the fact
that wave mechanics originated from a relativistic mechanics
where one just objetivizes bits of matter relative to an
observer, but leaves spacetime to always
play the role of coordinates labeling events that occur
through the emergence in spacetime of such bits of
matter. It is the author's contention that,
relative to an observer,
one would also need a relativistic theory of
momentum-energy itself in order to objetivize bits of the
spacetime -i.e. bits of space distances and time
durations, and hence derive the missing relations between
$R$ and $m$ and between $T$ and $m$,
following steps paralell to de Broglie's.
On the other hand, {\it a priori} presupposing that a
microscopic system is a particle or a wave would require
some appropriate physical conditions to be satisfied by
the system. 

In Einstein relativity space and time are relegated to play
the subjective role of elements of a language that is used
by the observer to describe his environment, and it is the
relation of the bit of matter (with its own space-time
trajectory) with the observer what makes the objective
reality out of which the world is constructed [5]. What
would be new in a relativistic formalism described in terms
of a momentum-energy continuum is the explicit 
renouncement to presuppose the Einsteinian objective
relation between the observer and bits of matter as a
necessarily establised and unique element of the possible
physical reality. Instead, we take all three notions, space,
time and matter -when considered independent of the observer-
as {\it a priori} being merely the elements of a subjective
language. 
The observer can then get related to either bits 
of matter or bits of space and time by some {\it introspective}
process that leads to either a distinct, purely theoretical
world picture, or to the design of related experiments and
observations, so that, depending on the very nature of the
system and the predisposition of the observer toward it,
either the bits of matter or the
bits of space and time become objectivized relative to
the observer, while space-time or momentum-energy remains
respectively 
relegated to play the subjective role of coordinates.

On the other hand,
in order to presuppose ``nothing''  about the
system an abstract
relativistic formalism should be established in which the
coordinate labeling events do not imply any
objetivization either of matter or of spacetime.
Consistently imposing then the appropriate physical
conditions on this formalism would finally result in
usual spacetime relativity or the alternate description
in terms of momentum-energy relativity
for objetivized bits of, respectively, matter or space
and time.
Quantities that one may take to play the role of the
coordinates labeling events in the generalized, abstract
formalism are the components of some unobjetivized action
$q^{\alpha}$, $\alpha=0,1,...,3$. Note that one can make
these coordinates simple dimensionless numbers by using
the Planck constant, thus showing the abstract character
of them. The usual line element of
Einstein relativity would then generalize to an action
element
\begin{equation}
ds^{(q)}=\left[(dq^{0})^{2}-
\sum_{i=1}^{3}(dq^{i})^{2}\right]^{\frac{1}{2}}.
\end{equation}
An inertial reference system for action coordinates $q^{\alpha}$
will then be an orthonormal frame, $q^{0},q^{1},q^{2},q^{3}$,
characterized by a constant value of the dimensionless
quantity $\frac{d{\bf q}}{dq^{0}}$. We assume (2.1) to be
relativistically invariant in any of such action reference
frames. Note 
however that since they do not correspond to visualizable
objetivized elements of the physical reality, the values
of these intervals cannot be measured by any experimental
devices. This abstract action interval should follow an
action line of the universe which at every point has a
tangent whose direction in action space is defined by a vector
with unit length given by
\begin{equation}
u_{(q)}^{\alpha}=\frac{dq^{\alpha}}{ds^{(q)}},
\end{equation}
with $u_{(q)}^{\alpha}u_{\alpha}^{(q)}=1$.

We regard the appropriate physical conditions that allow an
abstract wave-particle entity to be objetivized so that it
contains a bit of either space
and time (wave picture) or matter
(particle picture)
as being described by a mapping of the action coordinates onto
coordinates of, respectively, 4-momentum,
$dq^{\alpha}\rightarrow T_{0}cdp^{\alpha}$, and 4-position,
$dq^{\alpha}\rightarrow m_{0}cdx^{\alpha}$, where $T_{0}$
and $m_{0}$ are objetivized bits of time and matter, and
$c$ is the velocity of light.
In the first case, we allow the system to accommodate null
rays (null geodesics) along which  repetitive, reliable
measurements of its ``objetive'' spacetime characteristics
are enabled,
while the resulting unobjetivized 4-momentum components
$dp^{\alpha}$ are kept as coordinates that label events
with the above objetivized spacetime characteristics. In the
second case, the mapping allows the system to evolve 
along lines with
constant values of $\frac{d{\it p}}{dE}$ (which we call
{\it null cosmodesics}) and this permits repetitive, reliable
measurements of ``objective'' particle-like characteristics of the
system, while the resulting spacetime components $dx^{\alpha}$
are kept as coordinates that are used to label events with
the above objetivized particle-like characteristics.

The allowance of null cosmodesics to probe
the evolution of the
system makes then the action line element (2.1) and the
action velocity vector (2.2) to transform as
\begin{equation}
ds^{(q)}\rightarrow m_{0}cds^{(x)}
\end{equation}
\begin{equation}
u^{(q)\alpha}\rightarrow\frac{dx^{\alpha}}{ds^{(x)}}=u^{(x)\alpha},
\end{equation}
where $ds^{(x)}$ is the usual line element of spacetime Einstein
relativity and $u^{(x)\alpha}$ the corresponding velocity
of the universe. If we allow the system
to accommodate null geodesics in order to probe its evolution, 
then it is instead obtained
\begin{equation}
ds^{(q)}\rightarrow T_{0}cds^{(p)}
\end{equation}
\begin{equation}
u^{(q)\alpha}\rightarrow\frac{dp^{\alpha}}{ds^{(p)}}=u^{(p)\alpha},
\end{equation}
with $ds^{(p)}$ the line element in momentum-energy
coordinates and $u^{(p)\alpha}$ the velocity of the universe
defined on them. The invariance of the interval $ds^{(p)}$
would give rise to a formulation of relativity which is
formally equivalent to that of Einstein spacetime relativity
for inertial frames.

\subsection{Special relativity in momentum-energy}

In what follows I will formulate a momentum-energy
representation for relativity. In order for the
resulting theory to be self-consistent, such a
formulation should satisfy the following requirements.

(i) The kinematics of special relativity
(i.e. the relations between coordinate
labels) in the momentum-energy representation must
satisfy all mechanical Einstein four-momentum transformations,
and its associated mechanics (i.e. the quantities derived
from an action principle)
must in turn obey the usual
Lorentz transformations.

(ii) Whereas description of a given system in space-time
implies that such a system occupies just a space-time
part (often just a point)
from a necessarily larger system where at least
an {\it external} observer is also included, its description
in momentum-energy continuum requires
considering the system and the observer as located at
distinct particular values of momentum and energy intervals on the
same frame, so that no evolution of a system independent
of the observer is possible.

(iii) The nonrelativistic limit $c\rightarrow\infty$ of the
resulting mechanical relations between time durations
and space distances should produce either known or
rather trivial results, or not exist at all. 
The nonrelativistic
limit of the kinematical transformations of momentum
and energy must predict values of the energy
which depend on the chosen reference system, and
values of the momentum such that this behaved
as an absolute quantity.

The latter requirement needs some further explanation.
Consider a system {\it S} which evolves uniformly
(i.e. at a constant rate $\frac{d{\bf p}}{de}$) in the vacuum
momentum-energy continuum. Since, after requirement (i),
its evolution rate is $\frac{d{\bf p}}{de}=
\frac{{\bf v}}{c^{2}}$,
we can see why the components of momentum must become
absolute quantities in the nonrelativistic limit, where
energy will still depend on the bare velocity ${\bf v}$.
In such a limit, one would not expect the system {\it S}
with energy $e_{1}$ to interact with itseft with a
different energy $e_{2}$ because, then, the maximum rate
of signal propagation in momentum-energy, $\frac{1}{c}$,
becomes zero. 

Passing to the domain where $c$ is finite, we see that
the maximum rate of signal propagation in momentum-energy
is no longer zero and, therefore, the momentum components
become no longer absolute quantities. This will give rise
to the emergence of a purely relativistic interaction
of the system {\it S} with itseft when it evolves along
different values of the energy.
We can then introduce momentum-energy reference systems evolving
uniformly relative to each other with relative constant rates
$\frac{{\bf v}}{c^{2}}$,
so as an extended principle of relativity according to which all the laws
of nature are identical in all ``inertial'' momentum-energy reference
systems, if the equations expressing the laws and the events that
take place in such reference systems are all described in terms of
momenta and energies. Such laws must then be invariant with respect
to transformations of momenta and energies from one momentum-energy
reference system to another.

A differential interval defined in one of such reference systems
can be given by
\begin{equation}
ds^{(p)2}=\frac{de^{2}}{c^{2}}-dp_{x}^{2}-dp_{y}^{2}-dp_{z}^{2}.
\end{equation}
The principle of relativity for momentum-energy continuum implies
that $ds^{(p)}$ will be the same in all inertial momentum-energy systems,
and leads to the definition of a proper energy given by
\begin{equation}
de=\frac{de'}{\gamma},\;\;  \gamma
=\left(1-\frac{v^{2}}{c^{2}}\right)^{\frac{1}{2}}. 
\end{equation}

Let us consider two inertial momentum-energy reference
systems independently evolving with a relative rate
$\frac{{\bf v}}{c^{2}}$. From the above discussion it follows
that if the energy origin is chosen at the point where
both systems coincide, and such systems evolve so that
their $p_{x}$-axes always coincide, then we will have
in the limit $c\rightarrow\infty$
\begin{equation}
p_{x}=p_{x}',\; p_{y}=p_{y}',\; p_{z}=p_{z}',\; e=e'+p_{x}v .
\end{equation}
On the other hand, if $c$ is kept finite,
it is easy to see that the transformations that leave invariant the
interval are
\[p_{x}=\frac{p_{x}'+\frac{v}{c^{2}}e'}{\gamma},\;
p_{y}=p_{y}',\;\] 
\begin{equation}
p_{z}=p_{z}',\; e=\frac{p_{x}'v+e'}{\gamma}, 
\end{equation}
which, in turn, coincide with the transformation formulas 
for momentum-energy
4-vector of Einstein relativistic mechanics. Equations (2.10) 
lead to expressions
for the transformations of velocities, general 4-vectors, and unit
4-velocities, which exactly coincide with those of Einstein relativistic
kinematics, and reduce to (2.9) as $c\rightarrow\infty$. Thus, the
transformations (2.10) do satisfy the kinematical parts of the
requirements in (i) and (iii).

In order to formulate the relativistic mechanics in
momentum-energy representation,
let us consider a free system
evolving in the momentum-energy continuum. For such a system there
should exist a certain integral (the counterpart to action of
Einstein relativity in momentum-energy continuum) which has the
minimum value for actual evolution 
of the system in the momentum-energy continuum. This integral must
have the form
\begin{equation}
P=-\beta\int_{a}^{b}ds^{(p)}=
-\frac{\beta}{c}\int_{e_{1}}^{e_{2}}de\gamma =
\int_{e_{1}}^{e_{2}}\tilde{L}de,
\end{equation}
where $\int_{a}^{b}$ is an integral along a momentum-energy world line
of the system between two particular events characterizing the
momentum of the system when it has energies $e_{1}$ and $e_{2}$,
and $\beta$ is some constant that characterizes the system. The
coefficient $\tilde{L}$ of $de$ plays the role of a Lagrangian and
has the physical dimensions of a time. For $P$ to have the
dimensions of an action, unlike Einstein relativity
where each system is characterized by its rest energy $mc^{2}$, here
each system should be characterized by the complementary quantity
to its rest energy, that is its rest time $T_{0}$. We take therefore
$\beta=cT_{0}$, and hence the integral $P$ for a free
temporal system becomes
\begin{equation}
P=-T_{0}\int_{e_{1}}^{e_{2}}de\gamma ,
\end{equation}
with $\tilde{L}=-T_{0}\gamma$.

Instead of a momentum and an energy, the mechanical system will now
be described by a space distance $R$ and a time duration $T$.
Assuming the momentum-energy coordinate space to be homogeneous,
so that the properties of the system remain invariant under
infinitesimal parallel displacements of rate $\frac{{\bf v}}{c^{2}}$
and energy $e$, the quantities $R$ and $T$ would be conserved
and can be obtained using the same Lagrangian
principles as in classical mechanics, but in our complementary
representation, i.e.
\begin{equation}
R=\frac{\partial\tilde{L}}{\partial(\frac{v}{c^{2}})}=\frac{T_{0}v}{\gamma},
\; \; \; T=R\frac{v}{c^{2}}-\tilde{L}=\frac{T_{0}}{\gamma}.
\end{equation}

We have to check that the relativistic mechanics expressed by
(2.11)-(2.13)
is consistent with the full relativistic picture, i.e. we have to
check that by substituting space distance
and time duration given in (2.13), expressed as a
4-vector in terms of the corresponding 4-velocity, in the
transformation formulas for a general 4-vector, one obtains usual
Lorentz transformations. That this is indeed the case can be readily
seen by using the principle of least action, $\delta P=0$, and
$ds^{(p)}=(dp_{\alpha}dp^{\alpha})^{\frac{1}{2}}$, 
with $p^{0}=\frac{e}{c}$, $p^{1}
=p_{x}$, $p^{2}=p_{y}$, $p^{3}=p_{z}$. We then obtain $\delta P=
-T_{0}u^{(p)}_{\alpha}\delta p^{\alpha}$, where $u^{(p)}_{\alpha}=
\frac{dp^{\alpha}}{ds^{(p)}}=u^{(x)}_{\alpha}=u_{\alpha}$,
$u^{(x)}_{\alpha}$ being the Einstein unit 4-velocity (see the 
next subsection). It follows that
\begin{equation}
x_{\alpha}=-\frac{\partial P}{\partial p^{\alpha}}=(cT,R)=T_{0}u_{\alpha}
\end{equation}
is the distance 4-vector. It turns out that the square of the
length of momentum 4-vector, $(p^{0})^{2}-
\sum_{i=1}^{3}(p^{i})^{2}$,
is invariant under transformations (2.10). 
Generalizing to any 4-vector
$A^{\alpha}$ which transforms like the components of the momentum
4-vector under (2.10), we recover the usual transformation formulas
for 4-vectors of Einstein relativity. It is now inmediately seen
that by substituting (2.14) into such formulas, one obtains usual
Lorentz transformations. This completes fullfilment of
requirement (i).

We also note that the formula for $T$ in
(2.13) has no nonrelativistic counterpart. In fact, in the
limit $c\rightarrow\infty$, we obtain from (2.13)
\begin{equation}
T\approx T_{0}+\frac{T_{0}v^{2}}{2c^{2}}\approx T_{0},\;\; R\approx T_{0}v,
\end{equation}
i.e. the nonrelativistic limit of $T$ and $R$ reduces, respectively,
to the rest time and a distance-velocity law which may be
trivially interpreted as the customary definition of velocity.

On the other hand, it also follows from (2.13)
\begin{equation}
R=Tv
\end{equation}
\begin{equation}
T^{2}c^{2}=R^{2}+T_{0}^{2}c^{2}.
\end{equation}
Expression (2.16) should now correspond to 
the relativistic expression for
the definition of velocity of the object.
We finally note that (2.17) must correspond to the analogue of
the usual relativistic Hamiltonian in our complementary 
momentum-energy formalism
for relativity. If we express time $T$ 
in terms of the distance $R$, then
we have a complementary relativistic "Hamiltonian"
\begin{equation}
T\equiv H_{T}=\frac{1}{c}\left(R^{2}+T_{0}^{2}c^{2}\right)^{\frac{1}{2}},
\end{equation}
which has the physical dimension of a time. 
Law (2.18) must correspond
to the Minkowskian function $F$ which is the conjugate counterpart
to Hamiltonian and whose existence has been recently suggested
[6]. It describes the way in which objectivized bits of space
distance,$R$, and time interval, $T$, are related to each other.

We still have to check that our mechanical relation (2.17)
satisfies requirement (iii). Unlike the conventional
Hamiltonian of Einstein relativity, which in the limit
$c\rightarrow\infty$ produces the known nonrelativistic
Hamiltonian $\frac{p^2}{2m}$ plus the rest energy, the
relation (2.18) gives only the rest time $T_0$ in that limit
where, therefore, it induces no mechanical effects. Of course,
for high-velocity experiments one would expect time $T$ to
increase with velocity $v$ and $T_0$, such as it is also 
predicted by Einstein relativity and verified many times in
laboratory experiments.
When suitably generalized to noninertial frames so that it
becomes applicable to the whole universe, this law will describe the
cosmological evolution in the vacuum momentum-energy continuum
(see Sec. IIIB).

\subsection{The R-m and T-m relations}

We note now that the velocities of the universe $u^{(x)\alpha}$,
$u^{(p)\alpha}$ and $u^{(q)\alpha}$ are all the same; i.e.:
\begin{equation}
u_{\alpha}^{(x)}=u_{\alpha}^{(p)}=u_{\alpha}^{(q)}
=u_{\alpha}=\frac{v_{\alpha}}{c\gamma}
\end{equation}
This invariance would be a particular example of an invariance notion
which refers to quantities that preserve their values in all the
above three types of coordinate systems. Of course, all
dimensionless quantities that can be formed in the theory should
respect this kind of invariance which we hereafter refer to as
{\it representation invariance}. Thus, the de Broglie theorem
of phase harmony [2] can be regarded to be a consequence from this
invariance. We can in fact visualize any microscopic entity as
evolving along lines of the universe on three distinct sheets.
Evolution on the action sheet would describe an unobjetivized
wave-particle entity propagating with rate $\frac{d{\bf
q}}{dq^{0}}$ and having a pure action phase $\varphi^{q}$. On
the action line of the universe the entity would carry no definite
observable energy or characteristic time. The action sheet
can be unfolded by the above-mentioned mappings into the usual 
spacetime sheet and a momentum-energy sheet, each with the
corresponding line of the universe projected on it. Along the
spatial line of the universe on the spacetime sheet, the entity
would manifest as a bit of energy propagating 
on that sheet with given
velocity ${\bf v}$, and along the momentum line of the universe
on the momentum-energy sheet, it manifested like a bit of time
``propagating'' with corresponding rate $\frac{d{\bf p}}{de}=
\frac{{\bf v}}{c^{2}}$ in momentum-energy, or like the phase
wave with phase $\varphi^{(x)}$ and velocity $\frac{c^{2}}{{\bf
v}}$ , relative to the spacetime sheet. Likewise, projected on
the spacetime sheet, the entity would manifest like the phase
wave with phase $\varphi^{(p)}$ and propagation rate
$\frac{1}{{\bf v}}$, relative to the momentum-energy sheet.
Since the phase is dimensionless, we must then have
\begin{equation}
\varphi^{(q)}=\varphi^{(x)}=\varphi^{(p)}=\varphi .
\end{equation}
These equalities would in fact represent a generalization
of the de Broglie theorem of phase harmony.

Let us consider any two points $P$ and $Q$ along the action line
of the universe on the action sheet. We can then form the integral
\begin{equation}
-\int_{P}^{Q}ds^{(q)}=-\int_{P}^{Q}u_{\alpha}^{(q)}dq^{\alpha},
\end{equation}
which should have a stationary value. It is then possible to
introduce a general vector of the universe
\begin{equation}
J_{\alpha}=u_{\alpha} ,
\end{equation}
and a principle of least action such that
\[\delta\int_{P}^{Q}J_{\alpha}dq^{\alpha}\]
\begin{equation}
=m_{0}c\delta\int_{P'}^{Q'}J_{\alpha}dx^{\alpha}
=T_{0}c\delta\int_{P''}^{Q''}J_{\alpha}dp^{\alpha}=0,
\end{equation}
where $P'=\frac{P}{m_{0}c}$, $P''=\frac{P}{T_{0}c}$, and
similarly for $Q'$ and $Q''$.

On the other hand, rays of the universe will be determined by
the Fermat principle [2], i.e.
\begin{equation}
\delta\int_{P}^{Q}d\varphi = 0,
\end{equation}
for the representation-invariant phase $d\varphi$
\begin{equation}
d\varphi=2\pi O_{\alpha}^{(q)}dq^{\alpha}
=2\pi O_{\alpha}^{(x)}dx^{\alpha}=2\pi
O_{\alpha}^{(p)}dp^{\alpha}, 
\end{equation}
where the $O_{\alpha}$'s are the wave vector of the universe [2]
on
the respective representation.

If null cosmodesics are allowed to occur and be used as probes
to follow the evolution of the system, then
\begin{equation}
O_{\alpha}^{(q)}=\frac{O_{\alpha}^{(x)}}{m_{0}c},
\end{equation}
and if, alternatively, null geodesics are permitted to probe the
evolution of the system, we obtain
\begin{equation}
O_{\alpha}^{(q)}=\frac{O_{\alpha}^{(p)}}{T_{0}c} .
\end{equation}
The de Broglie's extension of the quantum relation [2] generalizes
then to read
\begin{equation}
hO_{\alpha}^{(q)}=u_{\alpha} ,
\end{equation}
with $h$ the Planck constant.
Thus, whereas (2.28) yields the known Einstein-de
Broglie relations between momentum and energy and, respectively,
wavelength and frequency whenever null cosmodesics are allowed
to occur and be used to follow the evolution of the system, 
as far as null geodesics are used to do that, (2.28) gives rise
to the new fundamental quantum relations
\begin{equation}
\mu=mc=\frac{h}{R}, \; \; T=h\Omega,
\end{equation}
where Eqn. (2.14) has been used, and $\Omega$ is the energy
frequency in momentum-energy continuum.

The first of relations (2.29) provides us with the wanted
relation between a discretized
$R$ and mass $m$. It allows the interpretation
of the wave function $\Psi(x)$ in $x$-representation as
the probability amplitude for a microscopic system to be a
particle with mass $m$ when one presupposes the system to
be a wave. One could say that a particle is not but just
a wave propagating in momentum-energy continuum with
characteristic ``wavemomentum'' $\mu$. The relation
$R\mu=h$ promotes the definition of the relativistic
Compton wavelength to the same fundamental status as that
played by the de Broglie relation $p\lambda=h$. This new
fundamental relation already has been therefore tested in
all those atomic-physics experiments aiming at e.g. measuring
the relativistic interaction between the electron and the
Coulomb field produced by the hydrogen nucleus, corresponding
to the fine-structure Darwin term. In particular, the first
of equations (2.29) would predict that an electron undergoing
Darwin interaction would be sensible to the ensemble of values
taken by the Coulomb field within a spatial domain which
would decrease as the electron is excited to upper energy
levels. 

Equivalently, the fourth-component relation $T=h\Omega$
provides us with the missing relation between time and
a particle-like property, and
discretizes an objectivized time $T$ for the system which
corresponds to the time scale that light waves (null geodesics)
would last in traversing the spatial domain $R$. According
to it, the objectivized
time appears to be quantized in
discrete portions, each carrying the total energy of the
system. This entails no violation of energy conservation
as the time portions are independent of each other.
Both relations in (2.29) have no counterpart in the
nonrelativistic limit $c\rightarrow\infty$.

\subsection{Wave mechanics in momentum-energy}

A quantum-mechanical wave equation can also be derived from (2.18)
by introducing the operators $\hat{T}=i\hbar\frac{\delta}{\delta e}$
and $\hat{R}=i\hbar\frac{\delta}{\delta p}$. Using a wave function
$\Psi\equiv\Psi(p,e)$, we obtain
\begin{equation}
-\hbar^{2}\frac{\partial^{2}\Psi}{\partial e^{2}}=
\frac{1}{c^{2}}\left(-\hbar^{2}\frac{\partial^{2}}{\partial 
p^{2}}+T_{0}^{2}c^{2}+V(p)\right)\Psi,
\end{equation}
where we have introduced a generic potential $V(p)$. This is the
counterpart in momentum-energy to the Klein-Gordon equation. If,
as it is the case for the whole universe, the system is closed, then
one would expect a discrete $T$-spectrum which would associate with
an infinite set of universes "frozen" at the given 
eigenvalues of $T$. This
spectrum would only tend to 
become continuous in the classical region that
corresponded to very large values of $T$. We finally note that
the quantum description of systems that show
time asymmetry could only be accounted for whenever we assume a
haft-integer intrinsic angular momentum for the whole system, so that,
instead of (2.30), one would have a Dirac-like wave equation
\begin{equation}
\left(\gamma^{\alpha}\frac{\partial}{\partial p^{i}}+cT_{0}+V(p)\right)\Psi(p)=0,
\end{equation}
with $\gamma^{\alpha}$ the $4\times 4$ Dirac matrices, which is invariant
under $e\rightarrow -e$, but not under $T\rightarrow -T$. Indeed,
just as for antimatter in momentum representation, the negative
time states could not be physically ignored, since there is nothing
to prevent a system from making a transition from a state of positive
time to a state of negative time. Equivalence between the two
relativistic quantum-mechanical representations manifests
here in the sense that states with negative time in momentum
representation should be equivalent to states with negative
energy in position representation as far as an antiparticle
moving forward in time is equivalent to the corresponding 
particle moving backward in time.

Actually, in Einstein relativity the Minkowskian coordinates $x^{0}=t$
and $x^{i}$ have a double function: they serve as labels for
the events but at the same time they also inform us through
the Lorentz transformations about actual time durations and
space distances, measurable with clocks and meters. Moreover,
although in Einstein relativity momentum components and
energy can never be taken to label real events, they can
be nevertheless obtained as actual quantities from the
associated relativistic mechanics where mass is introduced
as an objetivized bit of matter. Likewise, in the
momentum-energy representation of special relativity one
would expect the coordinates energy $p^{0}=e$ and
momentum $p^{i}$ to have also a double function: serving
as labels of events characterized by objetivized bits of
spatial sizes and time durations, and informing us about
the actual values of the energy and momentum of the 
system. Such values should be the same as those predicted
by Einstein relativistic mechanics. In momentum-energy
representation of relativity, one would also obtain the
same transformation formulas for time durations and space 
distances as in spacetime relativity, though in this case 
these quantities are given as mechanic rather than kinematic
quantities.

Since the Klein-Gordon relativistic wave equation gives the
eigenenergies of the system in terms of mass eigenvalues, 
$e_{n}=m_{n}c^{2}$, and time
periods are related to the corresponding wavelengths by
an explicit relation, one should expect the quantum theory
derived from relativity in momentum-energy coordinates to
be formulated in terms of wave functions which admits a
completely equivalent interpretation to that of the wave
functions of the quantum theory derived from Einstein
relativity, when both are applied to inertial systems.
Therefore, the nongravitational quantum theory formulated
above must be completely equivalent to that derived from
Einstein special relativity.

\subsection{The cosmological field}

The conclusion obtained in the precedent subsection is no
longer valid for noninertial frames. In general relativity
we must distinguish between coordinate labels from proper
intervals, entering at the two totally different levels
that correspond, respectively, to differential topology
and metric geometry. In spacetime general relativity showings
of a physical clock are predicted not only by the labels
that distinguish events, but also by the metric, and the
change of the metric respect to spacetime coordinates
describes at the same time a dynamical quantity: the
gravitational field. In general theory of relativity
formulated in terms of momentum-energy coordinates, besides
mechanical time durations and space distances, one would
likewise expect the emergence of a new quantity: the metric
of the momentum-energy continuum, $f_{\alpha\beta}\equiv
f_{\alpha\beta}(p^{\iota})$, which would help, together with the
$p^{\iota}$-coordinate labels, to construct actual
momentum and energy intervals.

Although as far as it describes the geometry of the
momentum-energy continuum, the dimensionless metric tensor 
$f_{\alpha\beta}$
would be the same as the usual tensor $g_{\alpha\beta}$ 
by representation
invariance, its variations with respect to momentum-energy
coordinates should, at the same time, describe an independent
quantity with ``dynamical'' content by itself; i.e.: a new
field which would generally
differ from gravity and only coincides with 
this under particular, limiting conditions. These two fields
would in general induce different behaviours in systems acted
upon by them.

It is in this sense that curved spacetime and curved
momentum-energy are not equivalent representations of a
unique general-relativity theory. I will give now some
arguments in support of the interpretation that the
variations of the momentum-energy metric 
$f_{\alpha\beta}(p^{\iota})$
with respect to the momentum-energy coordinates must
describe cosmological interactions.

\noindent (i) Because the dimensionless quantities 
$f_{\alpha\beta}$,
as regarded as the components of a metrical tensor, are
representation invariant, we should have 
$f_{\alpha\beta}=g_{\alpha\beta}$
and therefore, if $f_{\alpha\beta}$ is 
taken to describe a cosmological
field, appropriate solutions of the usual Einstein equations
satisfying Weyl and cosmological principles are also
cosmological solutions.

\noindent (ii) For the reasons discussed in the precedent
subsections, one would not expect a cosmological field to
have nonrelativistic counterpart. This must actually be the case
for the interactions described by the field equations derived
from $f_{\alpha\beta}$. Nevertheless, one can consider the limit
of very small but yet nonzero values of $\frac{v^{2}}{c^{2}}$,
where a very weak but still nonzero cosmological field with
potential $\rho$ is present. Assuming this field
to be described from the metric $f_{\alpha\beta}$,
$\rho\equiv\rho(p^{\iota})$, the time-Lagrangian in
momentum space could be written as
\begin{equation}
\tilde{L}=-T+\frac{Tv^{2}}{2c^{2}}-Tc^{2}\rho ,
\end{equation}
where, similarly to as the nonrelativistic gravitational 
potential goes like $(\frac{dr}{dt})^{2}$, i.e. like
a squared velocity, the potential $\rho$ should go 
like
$(\frac{dp}{de})^{2}=(\frac{dr}{c^{2}dt})^{2}$, i.e. like
the inverse of a squared velocity. Hence, in the
nonrelativistic limit $\frac{v}{c}\rightarrow 0$, $\rho$
will in fact strictly vanish. The situation we shall
nevertheless consider is one where
\begin{equation}
0 < c^{2}\rho << 1 .
\end{equation}
Comparing in this case the action derived from (2.32),
$P=\int\tilde{L}de$, with the general expression for action in
momentum-energy relativity, $P=-Tc\int ds^{(p)}$, we get
\[ds^{(p)2}=\left(\frac{1}{c^{2}}+\frac{v^{4}}{4c^{6}}
+c^{2}\rho^{2}-\frac{v^{2}}{c^{4}}+2\rho
-\frac{v^{2}\rho}{c^{2}}\right)de^{2} .\]
Taking into account that $dp=\frac{vde}{c^{2}}$ and
$dp^{0}=\frac{de}{c}$, we obtain then for very small
$\frac{v}{c}$ 
\begin{equation}
f_{00}\simeq 1+2c^{2}\rho, \; f_{0i}=0, \;
f_{ii}\simeq -(1+c^{2}\rho) ,
\end{equation}
where only the second and third terms in the r.h.s. of the
expression for $ds^{(p)2}$ above have been disregarded.

On the other hand, by analogy with the Einstein equations,
the field equations in momentum-energy
representation can be written
\begin{equation}
R(f,p)_{\alpha\beta}=4\pi K\left(S_{\alpha\beta}
-\frac{1}{2}f_{\alpha\beta}S\right),
\end{equation}
where $R(f,p)_{\alpha\beta}$ is the Ricci tensor espressed in terms
of tensor $f_{\alpha\beta}$ and coordinates 
$p^{\gamma}$, and $K$ is the
coupling constant for the new field. Since this constant has the
dimension of a conventional force, one can regard it
as the universal force exerted upon the system by a
universal constant field other than $\rho$, which is defined in
spacetime. The sole field which appears to be able to
generate such a force is gravity and since this is
attractive, $K$ must be negative: $K=-\mid K\mid$. Finally,
the tensor $S_{\alpha\beta}$ is a {\it space-time} 4-tensor, the
counterpart of the momentum-energy 4-tensor of Einstein
equations in momentum-energy relativity. When the involved
velocities are small compared to the velocity of light, we
have
\begin{equation}
S_{\alpha}^{\beta}\simeq u_{\alpha}u^{\beta}\tau_{0} ,
\end{equation}
so that the dominating term in this tensor becomes $S_{0}^{0}
\simeq\tau_{0}$. The parameter $\tau_{0}=\frac{T}{V_{p}}$, 
where $V_{p}$ is the 3-volume in momentum space, accounts for
the time that characterizes the system in momeentum-energy
continuum in a unit momentum volume. Then, Eqn. (2.35)
reduces to
\begin{equation}
R(f,p)_{0}^{0}\simeq -4\pi\mid K\mid\tau_{0} .
\end{equation}
We note that the terms in (2.37) which contain derivatives
of the affine connections in momentum-energy,
$\Gamma(f,p)^{\alpha}_{\beta\gamma}$, 
with respect to $\frac{e}{c}$ involve
extra power $c$ and therefore are large as compared to the
derivative with respect to the momenta $p^{i}$,
$i=1,2,3$. Hence, we can approximate
\begin{equation}
\partial_{e}\left(\frac{1}{2}f^{ik}\partial_{e}f_{ik}\right)
\simeq 4\pi\mid K\mid\tau_{0} .
\end{equation}

From (2.34) and (2.38) we obtain
\begin{equation}
\ddot{\rho}\simeq 4\pi\mid K\mid\tau_{0},
\end{equation}
in which an overhead dot means derivative with respect to the
energy-coordinate $e$. Direct integration of (2.39) yields
\begin{equation}
\rho(e)\simeq 4\pi\mid K\mid\tau_{0}e^{2}+K_{1}e+K_{2},
\end{equation}
where the integration constants $K_{1}$ and $K_{2}$ should
be both zero since $\rho$ must vanish at the sourceless limit
$\tau_{0}\rightarrow 0$.

Let us now consider the more important case of a constant
$\rho$-field, meaning by that a field $\rho$ which does not
depend on $p^{0}$. In this case, we obtain from (2.33) and
(2.37)
\begin{equation}
c^{2}\bigtriangleup_{p}\rho\simeq -4\pi\mid K\mid\tau_{0} ,
\end{equation}
where $\bigtriangleup_{p}=\frac{\partial^{2}}{\partial 
p_{i}\partial p^{i}}$. Eqn. (2.41) is formally
the same as the Newtonian Poisson equation of 
nonrelativistic gravity, except for the sign in the r.h.s.
The latter feature shows the essentially repulsive character
of field $\rho$ which, therefore, could be a good candidate
to describe cosmological interactions. The analogy between
(2.41) and the Poisson equation for gravity allows one to
solve (2.41) in a way which parallels the solution of the
Coulomb law. In the simplest situation of a single system
with characteristic time $T$, we have
\begin{equation}
\rho\equiv\rho(p^{i})\simeq\mid
K\mid\int\frac{\tau_{0}dV_{p}}{c^{2}P}=\frac{\mid K\mid
T}{c^{2}P} .
\end{equation}
We define the
{\it p-force} between two sources $T$ and $T'$ as
\begin{equation}
F_{p}=-c^{2}T'\frac{\partial\rho}{\partial P}\simeq\frac{\mid
K\mid TT'}{P^{2}}.
\end{equation}
It is worth noticing that the p-force $F_{p}$ has the dimension
of the inverse of a conventional force, that is the dimension
of the Newton constant $\frac{G_{N}}{c^{4}}$. Therefore, we
can regard $G_{N}$ as the quantity that characterizes the
constant $\rho$-field in our universe. Since $F_{p}$ is
repulsive $G_{N}$ is then positive. On the other hand, since
$\rho$ does not depend on the position $x^{i}$, it must
take on the same value ($\rho_{0}$ say) at any two distant
spatial points in a system, provided these points are
characterized by the same momenta and times.
Assuming then that the
total mass of the system is $M_{0}$ and taking $l=cT$,
$V=\frac{P}{M_{0}}$, we obtain from (2.42)
\begin{equation}
V=\left(\frac{\mid K\mid}{\rho_{0}M_{0}c^{3}}\right)l=Hl.
\end{equation}
Thus, $H$ can be interpreted as a Hubble constant and (2.44)
as a cosmological law.

\noindent (iii) For a constant $\rho$-field, $p^{0}$ should
be related to the proper energy $e$ by $e=\sqrt{f_{00}}p^{0}$.
In the case of weak $f_{\alpha\beta}$-fields, $e\simeq
cp^{0}(1+c^{2}\rho)$. Let us consider then the propagation of
a light ray in momentum-energy continuum when a weak constant
$\rho$-field is present. The light ray will be characterized
by an energy frequency $\Omega$ which would be given by the
derivative with respect to the energy coordinate of the 
phase-eikonal in momentum-energy, $\eta=-r_{\alpha}p^{\alpha}+\phi$
(with $\phi$ an arbitrary constant), for a ``plane wave''
$\sim e^{i\eta}$ in momentum-energy. As expressed in terms of
the energy $p^{0}$, the energy frequency becomes
$c\Omega_{0}=-\frac{\partial\eta}{\partial p^{0}}$,
and if we express it in terms of the proper energy $e$, we have
\begin{equation}
c\Omega=-c\frac{\partial\eta}{\partial e}
=-\frac{1}{\sqrt{f_{00}}}\frac{\partial\eta}{\partial p^{0}}
\simeq\frac{\Omega_{0}c}{1+c^{2}\rho} .
\end{equation}
We lift then the above restriction that spatial points of a
system have all the same local momenta and hence the same
values of field $\rho$. Thus, if a ray of light is emitted
at a point where the potential is $\rho_{1}$ and the energy
frequency is $\Omega$, then upon arriving at a point where
the potential is $\rho_{2}$ it will have an energy frequency
$\Omega\frac{1+c^{2}\rho_{1}}{1+c^{2}\rho_{2}}$. For an observer
at the arrival point the energy frequency would then be shifted
by an amount
$\bigtriangleup\Omega=\Omega\frac{c^{2}(\rho_{1}-\rho_{2})}{1
+c^{2}\rho_{2}}$ that corresponds to a proper-energy shift
given by
\begin{equation}
\bigtriangleup E\simeq Ec^{2}\left(\rho_{2}-\rho_{1}\right) ,
\end{equation}
where $E$ is the proper energy at the emission point where the
potential is $\rho_{1}$.

If we assume that the system is our universe and that every
point considered represents a galaxy of approximately the same
size and luminosity, then the light coming to our galaxy from
the inner regions of any other galaxy would be produced in
a physical environment similar to our own. In this case,
$\rho_{1}\simeq\rho_{2}$ and hence $\bigtriangleup E\simeq 0$.
However, as the light source separates from the core and enters
outer regions of the emitting galaxy where the momenta become
smaller, $\rho_{1}>\rho_{2}$ and from (2.46) $\bigtriangleup E <
0$. Although the approximation used in (2.46) breaks down as
$\rho_{2}$ increases, the above discussion appears to point
out that, rather than attributing this actually observed effect [7]
to the presence of some sort of dark matter, it would instead
be atributted to the noninvariance of proper energy under
propagation in curved momentum-energy.

In what follows, the above results will be taken to imply that
the field derived from variations of the metric tensor 
$f_{\alpha\beta}$
with respect to coordinates $p^{\iota}$ essentially describes
cosmological interactions. We shall therefore refer to this
field as the cosmological field.

\section{\bf Extended geometrodynamics}
\setcounter{equation}{0}

Let us introduce an arbitrary system of coordinates $X^{\iota}$
in a Riemannian spacetime, and an arbitrary system of
coordinates $P^{\iota}$ in a Riemannian momentum-energy.
Describe then a hypersurface in spacetime and a hypersurface
in momentum-energy by giving four functions $X^{\iota}(q^{i})$
of three action coordinates $q^{i}$ and four functions
$P^{\iota}(q^{i})$ of the same action coordinates $q^{i}$,
respectively; i.e.:
\begin{equation}
X^{\iota}=X^{\iota}(q^{i}), \; \; P^{\iota}=P^{\iota}(q^{i}) ,
\end{equation}
with $\iota=0,1,2,3$ and $i=1,2,3$; the 0-component in
momentum-energy corresponds to energy.
These two hypersurfaces are
thus {\it labeled} hypersurfaces [8], i.e.: in this case
two hypersurfaces
together with a common intrinsic action coordinate system
$q^{i}$ for them. Expressions (3.1) tell us that the point
of the $X^{\iota}(P^{\iota})$-hypersurface carrying the intrinsic
label $q^{i}$ is located in spacetime (momentum-energy) at the
point carrying the spacetime (momentum-energy) label
$X^{\iota}(P^{\iota})$. This implements
the unfolding discussed in Sec. II in the geometrodynamical
formalism (see Fig. 1).

\subsection{Deformations and relabelings}

Changes in a labeled hypersurface on a given projected
sheet (spacetime or momentum-energy) will generally induce
changes in the labeled hypersurface on the other projected
sheet. These labeled hypersurfaces are changed either by leaving
both fixed in the respective embedding spaces (spacetime and
momentum-energy) but relabeling uniquely their points, or by
deforming both hypersurfaces into other pair of hypersurfaces,
while leaving their labeling fixed. Any arbitrary change of
a pair of such hypersurfaces may be decomposed into these two
changes [8].

The first kind of changes represents a pure deformation of the
hypersurface in the Riemannian space without changing of labeling.
It can be carried out as follows. Start from hypersurfaces
$X^{\iota}(q^{i})$ and $P^{\iota}(q^{i})$. Draw geodesics
perpendicular to $X^{\iota}(q^{i})$ and cosmodesics
perpendicular to $P^{\iota}(q^{i})$. Move then along the
geodesic and cosmodesic that start from the point $q^{i}$,
eventually meeting a point of the deformed hypersurface
$\bar{X^{\iota}}$ and a point on the deformed hypersurface
$\bar{P^{\iota}}$, respectively. Attach to these points
the same label $q^{i}$ as that of the starting points, and
describe the displacement of $\bar{X^{\iota}}$ with respect
to $X^{\iota}$ by giving the proper time $\tau(q^{i})$
measured along the geodesic, and that of $\bar{P^{\iota}}$
with respect to $P^{\iota}$ by the proper energy
$\varepsilon(q^{i})$ measured along the cosmodesic. 
Repeating this operation at each point of the two original
hypersurfaces will give rise to two single functions
$\tau(q^{i})$ and $\varepsilon(q^{i})$ that describe the
operations of pure deformation of the two surfaces; i.e.:
$\varrho[\tau(q^{i})]$ and $\varrho[\varepsilon(q^{i})]$
(see Fig. 2).

Since in the curvilinear formalism
the cosmodesic does not match the respective geodesic,
the label of the end point on $\bar{X^{\iota}}(q^{i})$ will
not coincide with the corresponding label of the end point on
$\bar{P^{\iota}}(q^{i'})$, and therefore
the sheet $\Pi_{E}$
(or the sheet $\Pi_{T}$) of Fig. 2 should be deformed in 
an amount that allows these two final labels to exactly
coincide. But deforming e.g. the sheet $\Pi_{E}$ induces
an additional deformation of hypersurface $X^{\iota}(q^{i})$
itself. Hence, the action of an infinitesimal deformation
of hypersurface $X^{\iota}(q^{i})$ will be given by
\[\varrho_{T}[\tau(q)]X^{\iota}(q^{i})
\equiv\varrho_{E}[\delta N_{T}(q^{i}),
\delta N_{E}(q^{i})]X^{\iota}(q^{i})\]
\[=X^{\iota}(q^{i})\]
\[+{\bf n}^{\iota}(q^{i})\left[\delta
N_{T}(q^{i}) +\left(\frac{\partial N_{T}}{\partial
N_{E}}\right)(q^{i})\delta N_{E}(q^{i})\right]\]
\begin{equation}
=X^{\iota}(q^{i})+{\bf n}^{\iota}(q^{i})\delta N_{X}(q^{i}),
\end{equation}
where $\delta N_{T}$ and $\delta N_{E}$ account for the proper
time and the proper energy, respectively, ${\bf
n}^{\iota}(q^{i})$ is the unit normal to the hypersurface
$X^{\iota}$, and $N_{X}(q^{i})$ is a generalized lapse function
having the dimension of a time and is given by
\begin{equation}
N_{X}(q^{i})=N_{T}(q^{i})+\frac{\Omega}{\nu}N_{E}(q^{i}),
\end{equation}
in which we have used $\frac{\partial N_{T}}{\partial N_{E}}
=\frac{\Omega}{\nu}$, with $\Omega$ as given by the
second of expressions (2.29)
and $\nu$ is the usual frequency defined by the Einstein-de
Broglie relation $E=h\nu$.
Had we deformed sheet $\Pi_{T}$, instead of $\Pi_{E}$, then
we had obtained the infinitesimal deformation
\begin{equation}
\varrho_{E}[\varepsilon(q)]P^{\iota}(q^{i})
=P^{\iota}(q^{i})+{\bf m}^{\iota}(q^{i})\delta N_{P}(q^{i}) ,
\end{equation}
where ${\bf m}^{\iota}(q^{i})$ is the unit normal to the hypersurface
$P^{\iota}(q^{i})$, and
\[N_{P}(q^{i})=N_{E}(q^{i})+\frac{\nu}{\Omega}N_{T}(q^{i}) .\]

Similarly, relabeling is the operation (which we denote by
$\varrho[\bar{q}^{i}(q^{k})]$) that takes the label $q^{k}$
from fixed spacetime and momentum-energy points $X^{\iota}$
and $P^{\iota}$ and re-attaches it to the points
$\bar{X}^{\iota}$ and $\bar{P}^{\iota}$ which originally
had the label $\bar{q}^{i}(q^{k})$. Here deformations of
the spacetime (or momentum-energy) sheet are again necessary.
We have
\[\varrho_{E}[\bar{q}^{i}(q^{k})]X^{\iota}(q^{k})\]
\[=\varrho_{T}[\delta N_{T}^{i}(q^{k}),\delta
N_{E}^{i}(q^{k})]X^{\iota}(q^{k})
=X^{\iota}(q^{k})\]
\[+X^{\iota}(q^{k})_{,i}(q^{k})\left[\delta
N_{T}^{i}(q^{k})
+\left(\frac{\partial N_{T}^{i}}{\partial
N_{E}^{i}}\right)(q^{k})\delta N_{E}^{i}(q^{k})\right]\]
\begin{equation}
=X^{\iota}(q^{k})+X^{\iota}_{,i}(q^{k})\delta N_{X}^{i}(q^{k}) ,
\end{equation}
with the subscript $,i$ meaning the derivative with respect
to $q^{i}$, and we have used a
generalized shift function which is given by
\begin{equation}
N_{X}^{i}(q^{k})=N_{T}^{i}(q^{k})+\frac{\lambda}{\mu}N_{E}^{i}(q^{k}),
\end{equation}
where use of the de Broglie relation and the first of
expressions (2.29) has been made.

It is also obtained
\begin{equation}
\varrho_{E}[\bar{q}^{i}(q^{k})]P^{\iota}(q^{k})
=P^{\iota}+P^{\iota}_{,i}(q^{k})\delta N_{P}^{i}(q^{k}),
\end{equation}
in which
\[N_{P}^{i}(q^{k})=N_{E}^{i}(q^{k})+\frac{\mu}{\lambda}N_{T}^{i}(q^{k})
.\] 
Note that $\delta N_{X}$, as defined from (3.3), will give the
actual proper time separation, $T(q^{i})$ say, between any two
hypersurfaces and is generally different from $\tau(q^{i})$.
The set of deformations of hypersurfaces turns out to be an
infinitely dimensional set [8]  whose elements are characterized by
functions $T(q^{k})$, $\bar{q}^{i}(q^{k})$. One can define
generators for the relabeling. Let us use the notation
such that e.g. ${\cal H}_{i}(q^{k})\equiv{\cal H}_{iq}$,
$N_{X}^{i}(q^{k})\equiv N_{X}^{iq}$, etc. Then if the action of group
on the function space is expressed as
\begin{equation}
\varrho_{E}[N_{X}^{iq}]X^{\iota q'}=\bar{X}^{\iota q'}[X^{\kappa
q''},N_{X}^{iq}] ,
\end{equation}
then the generators can be identified through the infinitesimal
transformation
\[\varrho_{E}(\delta N_{X}^{iq})X^{\iota q'}\]
\begin{equation}
=X^{\iota q'}+\left.\frac{\delta\bar{X}^{\iota q'}[X^{\kappa
q''},N_{X}^{iq}]}{\delta N_{X}^{iq}}\right 
|_{N_{X}^{iq}=0}\delta N_{X}^{iq} ,
\end{equation}
in the neighborhood of the identity $N_{X}^{iq}=0$.

If we denote the coefficient for $\delta N_{X}^{iq}$ in (3.9) by
$\xi_{iq}^{iq'}$, the operators
\begin{equation}
X_{iq}=\xi_{iq}^{iq'}\frac{\delta}{\delta X^{\iota q'}}
\end{equation}
will be the generators of the relabelings. The vectors
$\xi_{iq}^{iq'}$ are obtained by comparing (3.9) with (3.5).
It follows
\begin{equation}
\xi_{iq}^{iq'}=X^{\iota}_{,i}(q')\delta(q,q') ,
\end{equation}
so that the generators of relabeling are
\begin{equation}
X_{iq}=X^{\iota}_{,i}(q)\frac{\delta}{\delta X^{\iota}(q)} .
\end{equation}
Proceeding similarly, we can also identify the generators of
pure deformations. They are:
\begin{equation}
X_{q}={\bf n}^{\iota}(q)\frac{\delta}{\delta X^{\iota}(q)} .
\end{equation}

The structure constants of the infinitely dimensional group
corresponding to relabelings and deformations are determined
from the commutation relations of their generators (3.12)
and (3.13). Of most interest is the commutator between two
generators (3.13)
\begin{equation}
[X_{q},X_{q'}]=-{\bf n}^{\kappa}(q')\frac{\delta{\bf
n}^{\iota}(q)}{\delta X^{\kappa}(q')}\frac{\delta}{\delta
X^{\iota}(q)}+(q,q') ,
\end{equation}
where $(q,q')$ means the same expression with $q$ and $q'$
interchanged. This antisymmetrization kills all terms in
$\frac{\delta {\bf n}^{\kappa}(q)}{\delta X^{\iota}(q')}$
which are proportional to the delta function $\delta(q,q')$
and, therefore, only the ``tilting'' term of
$\frac{\delta {\bf n}^{\kappa}(q)}{\delta X^{\iota}(q')}$
remains to contribute [8]. Such a tilting term has in this
case the form
\begin{equation}
-X^{i\iota}{\bf n}_{\iota}\left(\delta X^{\iota}_{,i}
+\frac{\lambda}{\mu}\delta P^{\iota}_{,i}\right) ,
\end{equation}
where the first term gives the change of ${\bf n}^{\iota}$
when the hypersurface $X^{\kappa}$ is displaced directly
by pure $X$-deforming by an amount $\delta X^{\kappa}(q)$,
and the second term accounts for the change of ${\bf n}^{\iota}$
produced by the displacement of hypersurface $X^{\kappa}$
induced by displacing $P^{\kappa}$ by an amount $\delta
P^{\kappa}(q)$. In (3.15) the Greek indices are raised and
lowered by $g_{\alpha\beta}$ (first term) and $f_{\alpha\beta}$
(second term), and the Latin indices by, respectively,
the metric tensors
\begin{equation}
g_{ik}= ^{4}g_{\iota\kappa}X^{\iota}_{i}X^{\kappa}_{k}, \; \;
f_{ik}= ^{4}f_{\iota\kappa}P^{\iota}_{i}P^{\kappa}_{k} ,
\end{equation}
where $X^{\iota}_{i}\equiv X^{\iota}_{,i}$ and
$P^{\iota}_{i}\equiv P^{\iota}_{,i}$. The terms (3.15) then
contribute by an amount
\[\frac{\delta_{\perp}{\bf n}^{\iota}(q)}{\delta X^{\kappa}(q')}
=-X^{i\iota}(q).{\bf n}_{\lambda}(q)\left(\frac{\delta
X^{\lambda}_{i}(q)}{\delta X^{\kappa}(q')}
+\frac{\lambda}{\mu}\frac{\delta P^{\lambda}_{i}(q)}{\delta
P^{\kappa}(q')}\right) \]
\[=-X^{i\iota}(q){\bf
n}_{\lambda}(q)\left(\delta^{\lambda}_{\kappa}\delta_{,i}(q,q')  
+\delta^{\lambda}_{\kappa}\delta_{,i}(q,q')\right)\]
\begin{equation}
-(q,q')
\end{equation}
\begin{equation}
\equiv -X^{i\iota}(q){\bf n}_{\kappa}(q)\delta_{,i}(q,q')-(q,q')
,
\end{equation}
in which the indices in the first term of (3.17) are raised
and lowered by $g_{\alpha\beta}$, $g_{ik}$ and those of the
second term in the same equation and in (3.18) by
$f_{\alpha\beta}$, $f_{ik}$ and also $g_{\alpha\beta}$,
$g_{ik}$. Substituting (3.18) in (3.14), we obtain
\[[X_{q},X_{q'}]\]
\begin{equation}
={\bf n}^{\kappa}(q'){\bf
n}_{\kappa}(q)X^{i\iota}(q)\delta_{,i}(q,q')\frac{\delta}{\delta
X^{\iota}(q)}-(q,q') .
\end{equation}
By employing then the usual procedure [8], we finally get
commutators with exactly the same formal structure as in
conventional geometrodynamics, but with the indices in the
r.h.s. being raised and lowered by $g_{ik}$ and also by
$f_{ik}$, which are defined in (3.16).

\subsection{Hamiltonian formalism}

A minimal representation of this extended formulation of
geometrodynamics should use as canonical variables both
the metric tensor $g_{ik}$ and the metric tensor $f_{ik}$
as well as their respective conjugate momenta $\pi^{ik}$
and $\omega^{ik}$. Our task now is to find the superhamiltonian
${\cal H}$ and the supermomentum ${\cal H}_{i}$ which should
be constructed out of the above metric tensors and momenta,
while respecting the commutation relations of geometrodynamics,
with an action functional
\[S=\int
d^{4}q(\pi^{ik}_{q}\dot{g}_{qik}+\omega^{ik}_{q}\dot{f}_{qik}\]
\begin{equation}
-N_{Xq}{\cal H}_{q}-N_{Xq}^{i}{\cal H}_{iq}),
\end{equation}
where $\dot{}\equiv\frac{\delta}{\delta q^{0}}$ and $N_{X}$ and
$N_{X}^{i}$ are given by (3.3) and (3.6), respectively. The
Hamiltonian ${\cal H}$ that corresponds to this action
functional determines then the change, $\delta F$, of any
arbitrary function $F$ of the geometrodynamic variables
($g_{ik},f_{ik},\pi^{ik},\omega^{ik}$) induced by the
deformation $\delta N_{X}=N_{X}\delta q^{0}$, $\delta N_{X}^{i}
=N_{X}^{i}\delta q^{0}$ of the two hypersurfaces. Under such a
deformation
\begin{equation}
\delta F=[F,{\cal H}_{q'}\delta N_{X}^{q'}+{\cal H}_{lq'}\delta
N_{X}^{lq'}], 
\end{equation}
where $N_{X}^{q'}$ and $N_{Xl}^{q'}$ are given by (3.3) and
(3.6). Specializing to pure relabeling ($\delta N_{X}=0$),
\begin{equation}
\delta F=[F,{\cal H}_{lq'}\delta N_{X}^{lq'}],
\end{equation}
and taking into account that both $g_{ik}$ and $f_{ik}$ tranform
like tensors and both $\pi^{ik}$ and $\omega^{ik}$ do like
tensor densities of weight 1 under relabeling, so that the
respective changes are given by the Lie derivatives of a
tensor and a tensor density, we obtain a set of equalities, i.e.
\[[g_{ikq},{\cal H}_{lq'}\delta N_{X}^{lq'}]=\frac{\delta {\cal
H}_{lq'}}{\delta\pi^{ikq}}\delta N_{X}^{lq'}\]
\[=g_{ik,l}\delta N_{X}^{l}+g_{il}\delta
N_{X,k}^{l}+g_{lk}\delta N_{X,i}^{l} ,\]

\[[f_{ikq},{\cal H}_{lq'}\delta N_{X}^{lq'}]=\frac{\delta {\cal
H}_{lq'}}{\delta\omega^{ikq}}\delta N_{X}^{lq'}\]
\[=f_{ik,l}\delta N_{X}^{l}+f_{il}\delta
N_{X,k}^{l}+f_{lk}\delta N_{X,i}^{l} ,\]

\[[\pi^{ikq},{\cal H}_{lq'}\delta N_{X}^{lq'}]=\frac{\delta {\cal
H}_{lq'}}{\delta g_{ikq}}\delta N_{X}^{lq'}\]
\[=(\pi^{ik}\delta N_{X}^{l})_{,l}-\pi^{il}\delta
N_{X,l}^{k}-\pi^{lk}\delta N_{X,l}^{i} ,\]

\[[\omega^{ikq},{\cal H}_{lq'}\delta N_{X}^{lq'}]=\frac{\delta {\cal
H}_{lq'}}{\delta f_{ikq}}\delta N_{X}^{lq'}\]
\[=(\omega^{ik}\delta N_{X}^{l})_{,l}-\omega^{il}\delta
N_{X,l}^{k}-\omega^{lk}\delta N_{X,l}^{i} ,\]
whose unique solution reads:
\begin{equation}
{\cal
H}_{iq}=-2\left(g_{ik}\pi^{kl}_{|i}+f_{ik}\omega^{kl}_{|i}\right),
\end{equation}
where the subscript $|i$ means the corresponding covariant
derivative. All derivatives in (3.23) are taken with respect
to the action-like coordinates $q^{i}$. These coordinates
were however defined such that hypersurface $X^{\iota}(q^{i})$
would correspond to a constant value of $q^{0}$. Due to the
mutual complementary character of $X$ and $P$, $q^{i}$ may
either be given by $q^{i}=\mu x^{i}$, when it is projected
onto spacetime, or by $q^{i}=\lambda p^{i}$ if is is projected
onto momentum-energy. Therefore, the covariant derivatives in
(3.23) can be written 
\begin{equation}
\pi^{kl}_{|i}=\frac{\pi^{kl}_{|x^{i}}}{\mu},\; \;
\omega^{kl}_{|i}=\frac{\omega^{kl}_{|p^{i}}}{\lambda}
\end{equation}
It then follows
\[\pi{\cal
H}_{iq}=-2\left(q_{ik}\pi^{kl}_{|x^{i}}
+\frac{\mu}{\lambda}\omega^{kl}_{|p^{i}}\right)\]
\begin{equation}
={\cal H}^{T}_{i}(x)+\frac{\mu}{\lambda}{\cal
H}^{E}_{i}(p)\equiv {\cal H}_{iX}
\end{equation}

Using the same ansatz as in usual geometrodynamics [8], we can
similarly obtain the superHamiltonian
\[{\cal H}_{q}=G_{qiklm}\pi^{ik}_{q}\pi^{lm}_{q}
-\left(\sqrt{g}R\right)_{q}\]
\begin{equation}
+F_{qiklm}\omega^{ik}_{q}\omega^{lm}_{q}-
\left(\sqrt{f}C\right)_{q} ,
\end{equation}
where
\[G_{qiklm}\]
\begin{equation}
=\frac{1}{2\sqrt{g}}\left(g_{ilq}g_{kmq}
+g_{imq}g_{klq}-g_{ikq}g_{lmq}\right)
\end{equation}
is the metric on usual superspace and $F_{qiklm}$, which is
given by the same expression as (3.26), 
but with the $g$'s replaced
for the corresponding $f$'s, is the metric on 
the equivalent superspace
constructed from momentum-energy coordinates. Finally, $R$
and $C$ are the scalar curvatures in the respective 3-space.

The superHamiltonian ${\cal H}_{q}$ will correspond to the
operation ${\cal H}_{q}\equiv\frac{\delta}{\delta q^{0}}$.
Depending on which of the two subspaces it is projected
onto, ${\cal H}_{q}$ can be written either as
\begin{equation}
{\cal H}_{q}=\Omega\frac{\delta}{\delta\tau}=\Omega {\cal H}_{X}
\end{equation}
or as
\begin{equation}
{\cal H}_{q}=\nu\frac{\delta}{\delta\varepsilon}=\nu {\cal
H}_{P} .
\end{equation}
Therefore,
\[{\cal
H}_{q}=\Omega\left(G_{qiklm}\pi^{ik}_{X}\pi^{lm}_{X}-
\left(\sqrt{g}R\right)_{X}\right)\]
\[+\nu\left(F_{qiklm}\omega^{ik}_{P}\omega^{lm}_{P}-
\left(\sqrt{f}C\right)_{P}\right)\]
\begin{equation}
\equiv\Omega\left({\cal H}^{T}(x)+\frac{\nu}{\Omega}{\cal
H}^{E}(p)\right)=\Omega{\cal H}_{X} .
\end{equation}

Using (3.2), (3.5), (3.24) and (3.29) in action (3.20) we obtain
\[S_{X}\propto\int
d^{4}x(\pi_{X}^{ik}\dot{g}_{Xik}\]
\begin{equation}
+\omega_{P}^{ik}\dot{f}_{Pik}
-N_{X}{\cal H}_{X}-N_{X}^{i}{\cal H}_{Xi}) ,
\end{equation}
where $\dot{}=\frac{\delta}{\delta\tau}$ when it is over $g$ and
$\dot{}=\frac{\delta}{\delta\varepsilon}$ when it is over $f$.
From $\frac{\delta S_{X}}{\delta N_{X}}$ we obtain the new 
Hamiltonian constraint
\begin{equation}
{\cal H}^{T}+\frac{\nu}{\Omega}{\cal H}^{E}=0,
\end{equation}
with ${\cal H}^{T}$ and ${\cal H}^{E}$
the supeHamiltonians of geometrodynamics and cosmodynamics
which, separately, are no longer zero in the present formalism.

Of course, one could re-formulate the above canonical formalism
in terms of the
cosmological field rather than the gravitational field. We would
then derive an action functional
\[S_{P}\propto\int
d^{4}p(\pi_{X}^{ik}\dot{g}_{Xik}\]
\begin{equation}
+\omega_{P}^{ik}\dot{f}_{Pik}
-N_{P}{\cal H}_{P}-N_{P}^{i}{\cal H}_{Pi}) ,
\end{equation}
where $N_{P}$ and $N_{P}^{i}$ are as given in Sec. IIIA, and
\begin{equation}
H_{P}={\cal H}^{E}(p)+\frac{\Omega}{\nu}{\cal H}^{T}(x)
\end{equation}
\begin{equation}
H_{iP}={\cal H}_{i}^{E}(p)+\frac{\lambda}{\mu}{\cal
H}_{i}^{T}(x) .
\end{equation}
These are the basic equations for the canonical formulation
of the cosmological field which we may call {\it cosmodynamics}.
From $\frac{\delta S_{P}}{\delta N_{P}}$, we would
then obtain again (after multiplying by $\nu$ and dividing by
$\Omega$ the resulting expression) 
the constraint (3.32). 

Clearly, physical systems that show observable gravitational
effects are usually of large size (even astrophysical black
holes are remarkably large). Such systems will then be
characterized by small values of $\nu$ and rather huge values
of $\Omega$. Hence, using the constraint ${\cal H}^{T}=0$
for them becomes an excellent approximation. However, for 
primordial black holes or in the very early universe, one
would expect the quantum characteristics of the systems 
to be exactly
the opposite -i.e. such systems would have large $\nu$ and
small $\Omega$. In this case, it would be the 
cosmological Hamiltonian which became approximately constrained
so that ${\cal H}^{E}\simeq 0$. Therefore, one would also
expect this constraint rather than the usual one to contain almost
all the relevant dynamical information required to describe
the latest stages of black-hole evaporation or the earliest
stages of the evolution of the universe.

Finally, we note
that by independently varying any of the two above
action functionals with
respect to either metric $g_{ik}$ or metric $f_{ik}$, we would
respectively obtain [9] Einstein equations and the cosmological
field equations (2.35).

\section{\bf Quantization}
\setcounter{equation}{0}

All of the essential steps that we shall adopt in what follows
are not but hints and guesses as they concern the quantization
of the canonical formalism developed in Sec. III. To my 
knowledge, there is no other way to proceed with the 
quantization of any field, not even for
inertial systems. We start with the action functional obtained
in the previous section for a gravitating system, i.e.
\[S_{X}\propto\int
d^{4}x(\pi_{X}^{ik}\dot{g}_{Xik}\]
\begin{equation}
+\omega_{P}^{ik}\dot{f}_{Pik}
-N_{X}{\cal H}_{X}-N_{X}^{i}{\cal H}_{Xi}) ,
\end{equation}
which has the same form as that of parametrized field theories,
but contains the additional (second) term, and
differs in the specific form of the superHamiltonian and
supermomentum which, in (4.1), read
\[{\cal H}_{X}={\cal H}^{T}(x)+\frac{\nu}{\Omega}{\cal
H}^{E}(p),\;\;\] 
\begin{equation}
{\cal H}_{iX}={\cal
H}^{T}_{i}(x)+\frac{\mu}{\lambda}{\cal H}^{E}_{i}(p) ,
\end{equation}
both being equal to zero.

Instead of (4.1), one could use the action functional relative
to the momentum-energy sheet, $S_{P}$, which is given in terms
of the superquantities ${\cal H}_{P}$ and ${\cal H}_{iP}$. In
the form given by (4.1) and (4.2), our action is prepared to
be quantized just on spacetime. Spacetime quantization would
proceed by turning into operators the metric $g_{ik}(x)$, the
momentum $\pi^{ik}(x)$ and, as a consequence from the fact
that they are given in terms of $g_{ik}(x)$'s and
$\pi^{ik}(x)$'s, the quantities ${\cal H}^{T}$ and
${\cal H}^{T}_{i}$, as well as the quantities ${\cal
H}^{E}$ and ${\cal H}^{E}_{i}$ by themselves. The
resulting operators are assumed to 
satisfy the commutation relations
(in what follows we set $\hbar=c=G=K=1$)
\begin{equation}
[g_{ik}(x),\pi^{lm}(x')]=\frac{1}{2}i\left(\delta^{l}_{i}\delta^{m}_{k}
+\delta^{m}_{i}\delta^{l}_{k}\right)\delta(x,x')
\end{equation}
\begin{equation}
[g_{ik}(x),g_{lm}(x')]=[\pi^{ik}(x),\pi^{lm}(x')]=0
\end{equation}
\begin{equation}
[{\cal H}^{T}(x),{\cal H}^{E}(p')]=
-i\frac{\Omega}{\nu}g^{br}(x)\delta(x)_{,b}\delta(x,x')_{,r}
\end{equation}
\[[{\cal H}_{j}^{T}(x),{\cal H}^{Ei}(p')]\]
\begin{equation}
=-i\frac{\lambda}{\mu}g^{br}(x)\delta
(x)_{,b}\delta(x,x')_{,r}\delta^{i}_{j} ,
\end{equation}
where the subscript $,l$ means derivation with respect
to $x_{l}$.

In order to proceed with the quantization of the 
complementary momentum-energy
canonical formalism, we would start with (3.33)-(3.35) and
similarly turn into operators $f_{ik}(p)$, $\omega^{ik}(p)$ and
hence the quantities ${\cal H}^{E}$, ${\cal H}^{E}_{i}$,
as well as the quantities ${\cal H}^{T}$ and ${\cal
H}^{T}_{i}$ by themselves. The resulting operators would
then satisfy the commutation relations
\begin{equation}
[f_{ik}(p),\omega^{lm}(p')]=\frac{1}{2}i\left(\delta^{l}_{i}\delta^{m}_{k}
+\delta^{m}_{i}\delta^{l}_{k}\right)\delta(p,p')
\end{equation}
\begin{equation}
[f_{ik}(p),f_{lm}(p')]=[\omega^{ik}(p),\omega^{lm}(p')]=0
\end{equation}
\begin{equation}
[{\cal H}^{E}(p),{\cal H}^{T}(x')]=
-i\frac{\nu}{\Omega}f^{br}(p)\delta(p)_{,b}\delta(p,p')_{,r}
\end{equation}
\[[{\cal H}_{i}^{E}(p),{\cal H}^{Tj}(x')]\]
\begin{equation}
=-i\frac{\mu}{\lambda}f^{br}(p)\delta
(p)_{,b}\delta(p,p')_{,r}\delta^{i}_{j} ,
\end{equation}
where the subscript $_{,l}$ denotes now derivation with
respect to $p_{l}$, instead
of $x_{l}$.

\subsection{Spacetime quantization}

Here, we shall restrict ourselves to explicitly deal with quantization
in spacetime. 
We shall adopt the metric representation in which the state
functional $\Psi$ will become a functional of the 3-metric
$g_{ik}(x)$ and the quantities ${\cal H}^{E}(p)$ and
${\cal H}^{E}_{i}(p)$, in such a way that
\begin{equation}
\Psi\equiv\Psi\left[g_{ik},{\cal H}^{E},{\cal H}^{E}_{i}\right]
\end{equation}
can be interpreted as containing the information about
the showings of clocks and meters
among its arguments. Then [10]:

(i) the 3-momentum $\pi^{ik}(x)$ is replaced by the variational
derivative with respect to the metric $g_{ik}(x)$
\begin{equation}
\hat{\pi}^{ik}(x)=-i\frac{\delta}{\delta g_{ik}(x)},
\end{equation}
and (ii) the quantities ${\cal H}^{T}(x)$ and
${\cal H}^{T}_{i}(x)$ are replaced by the functional
derivatives with respect to ${\cal H}^{E}(p)$ and
${\cal H}^{E}_{j}(p)$, respectively, i.e.:
\begin{equation}
\hat{{\cal H}}^{T}(x)=-i\frac{\delta}{\delta{\cal
H}^{E}(p)},\; \;
\hat{{\cal H}}^{T}_{j}(x)=-i\frac{\delta}{\delta{\cal
H}^{jE}(p)}.
\end{equation}
Following this procedure, we substitute these operators into
the superHamiltonian and supermomentum in spacetime
representation, and impose the general constraints (4.2) as
restrictions on the state functional, that is
\[\hat{{\cal H}}^{T}\Psi\equiv
-G_{iklm}(x)\frac{\delta^{2}\Psi}{\delta g_{ik}(x)\delta
g_{lm}(x)}\]
\begin{equation}
+\sqrt{g(x)}R(x)\Psi=
-i\frac{\delta\Psi}{\delta{\cal H}^{E}(p)}
\end{equation}
\begin{equation}
\hat{{\cal H}}^{T}_{i}\Psi\equiv 
2i\left(\frac{\delta\Psi}{\delta g_{ik}(x)}\right)_{|k}
=-i\frac{\delta\Psi}{\delta{\cal H}^{E}_{i}(p)}.
\end{equation}

These equations should always be different of zero, unless
for systems of infinite size.
We have therefore {\it deconstrained} our wave equations, 
leaving them in a manifest Schr"dinger-like 
form. As in the
parametrized field theories, equation (4.15) implies that
the state functional is unchanged under relabeling of the
hypersurfaces. Indeed, 
by a relabeling of the hypersurface the metric
must change into
\[g_{ik}\rightarrow\bar{g}_{ik}=g_{ik}-\delta N_{Xi|k}-\delta
N_{Xk|i}\]
while, since
${\cal H}^{E}_{i}$ has the dimension of a spacetime distance,
it undergoes the transformation
\[{\cal H}^{E}_{i}\rightarrow\bar{{\cal H}}^{E}_{i}
={\cal H}^{E}_{i}+\delta N_{Xi} .\]
For the state functional to be kept unchanged, one should then
have
\[\int d^{3}x\left(2\frac{\delta\Psi}{\delta g_{ik}}\delta
N_{Xi|k}-\frac{\delta\Psi}{\delta{\cal H}^{P}_{i}}\delta
N_{Xi}\right)=0. \]
By integrating by parts the first of these integrals and
taking into account the arbitrariness of $\delta N_{Xi}$,
we recover in fact the supermomentum wave equation (4.15).
The state functional thus depends on the spatial geometry
${\cal G}^{S}$ and physical distances ${\cal D}$, but not
on the particular metric and position chosen to represent
it. Likewise, one can show [8] the invariance of (4.14) under
pure deformations of the hypersurfaces, so that now the wave
functional will also depend on a generic time ${\cal T}$,
but not on any of the particular moments that may be chosen
to represent it. Thus,
\begin{equation}
\Psi\equiv\Psi\left[{\cal G}^{S},{\cal D},{\cal T}\right].
\end{equation}
It follows that the proper domain of the state functional
is an extended superspace which, besides on the 3-geometry,
depends also on suitable distance and time concepts. The
specific mathematical characteristics of such an extended
superspace will be considered in a future publication.
We have in
this way succeeded in separating suitably defined space
and time concepts from the dynamical variables.

\subsection{Momentum-energy quantization}

By following a completely parallel procedure, we finally
obtain in the case of the cosmological field
\begin{equation}
\hat{\omega}^{ik}(p)=-i\frac{\delta}{\delta f_{ik}(p)},
\end{equation}
\begin{equation}
\hat{{\cal H}}^{E}(p)=-i\frac{\delta}{\delta{\cal
H}^{T}(x)},\; \;
\hat{{\cal H}}^{E}_{j}(p)=-i\frac{\delta}{\delta{\cal
H}^{jT}(x)}.
\end{equation}
\[\hat{{\cal H}}^{E}\Phi\equiv
-F_{iklm}(p)\frac{\delta^{2}\Phi}{\delta f_{ik}(x)\delta
f_{lm}(x)}\]
\begin{equation}
+\sqrt{f(p)}C(p)\Phi=
-i\frac{\delta\Phi}{\delta{\cal H}^{T}(x)}
\end{equation}
\begin{equation}
\hat{{\cal H}}^{E}_{i}\Phi\equiv 
2i\left(\frac{\delta\Phi}{\delta f_{ik}(p)}\right)_{|k}
=-i\frac{\delta\Phi}{\delta{\cal H}^{T}_{i}(x)},
\end{equation}
with
\begin{equation}
\Phi\equiv\Phi\left[{\cal G}^{M},{\cal M},{\cal E}\right],
\end{equation}
where the subscript $_{|k}$ now means covariant derivative
with respect to the metric of momentum-energy,
${\cal G}^{M}$ denotes the geometry of a 3-momentum
superspace, and ${\cal M}$ and ${\cal E}$ some 
concepts of generic momentum and energy, defined parallely
to as for generic space and time concepts in the case
of spacetime quantization. Eqns. (4.17)-(4.21) form up
the essentials of the formulation of what we may call
quantum cosmodynamics, with the first two ones being
different of zero always unless for systems of zero size.

\subsection{Consistent operator-ordering}

Let us now see how the operator-ordering problem which appears
in conventional geometrodynamics can be worked out in our
extended formalism. The problem can be expressed by using
the hermitian ordering that corresponds to the quantum operators
proposed by Anderson [11]. In our extended 
formalism of geometrodynamics, 
Anderson's ordering translates into
\[{\cal
H}_{iX}=\frac{1}{2}\left[g_{ik}\pi^{kl}_{|l}+\pi^{kl}_{|l}g_{ik}
+\frac{\mu}{\lambda}\left(f_{ik}\omega^{kl}_{|l}+
\omega^{kl}_{|l}f_{ik}\right)\right]\]
\[{\cal H}_{X}=\frac{\pi^{ik}}{\sqrt{g}}\left(g_{il}g_{ml}
+g_{im}g_{kl}-g_{ik}g_{lm}\right)\pi^{lm}-\sqrt{g}R\]
\[+\frac{\nu}{\Omega}(g,f;\pi ,\omega;R,C) ,\]
where $(g,f;\pi ,\omega;R,C)$ denotes the same expression as
in all the explicited terms but with the $g$'s, $\pi$'s and
$R$ replaced for, respectively, the $f$'s, $\omega$'s and $C$.
The ordering problem is manifested through the commutator
between Hamiltonian constraints. For the ordering chosen,
in the present case one can find
\[2i[{\cal H}_{X}(x),{\cal H}_{X}(x, )]
=\delta(x,x')_{,r}\left[g^{rs}(x){\cal H}^{T}_{s}(x)\right.\]
\[+{\cal H}^{T}_{s}(x)g^{rs}(x)+g^{rs}(x'){\cal H}^{T}_{s}(x')\]
\[\left.+{\cal
H}^{T}_{s}(x')g^{rs}(x')\right]+\frac{\nu^{2}}{\Omega^{2}}(x,p;g,f;{\cal
H}^{T},{\cal H}^{E})\]
\begin{equation}
+\frac{2\nu}{\Omega}\left([{\cal H}^{T}(x),{\cal
H}^{E}(p')]+[{\cal H}^{E}(p),{\cal
H}^{T}(x')]\right) .
\end{equation}
We can readily check that the troublesome terms (those that have
factors $g^{rs}$ or $f^{rs}$ occurring to the right of the
${\cal H}^{T}_{s}$ or ${\cal H}^{E}_{s}$ [12] in the second
and third lines of (4.22)) are all canceled
by the commutators mixing Hamiltonian in $x$ with that in $p$
in the last line of (4.22). Using then $\delta_{,r}(p,p')
=\frac{\Omega}{\nu}\delta_{,r}(x,x')$ and (4.5) and (4.9), we
finally obtain
\[[{\cal H}_{X}(x),{\cal H}_{X}(x')]
=-\frac{1}{2}i
\left(g^{rs}(x){\cal H}^{T}_{s}(x)\right.\]
\[\left.+g^{rs}(x'){\cal
H}^{T}_{s}(x')+\frac{\nu}{\Omega}(x,p;g,f;{\cal H}^{T}_{s},
{\cal H}^{E}_{s})\right)\delta_{,r}(x,x')\]
\begin{equation}
=-\frac{1}{2}i\delta_{,r}(x,x')\left({\cal H}^{r}_{X}(x)+
{\cal H}^{r}_{X}(x')\right) .
\end{equation} 
Thus, (4.23) must vanish weakly. Since in the covariant form
${\cal H}^{T}_{i}$ or ${\cal H}^{E}_{i}$: 
(1) the interchange of momenta and
coordinates only leads to terms with $\delta_{,i}(x,x')$
or $\delta_{,i}(p,p')$ which can be put equal to zero,
and (2) the commutators
\begin{equation}
[{\cal H}^{T}_{j}(x),{\cal H}^{E}_{i}(p')]
=-i\frac{\lambda}{\mu}\delta(x)_{,j}\delta(x,x')_{,i}
\end{equation}
\begin{equation}
[{\cal H}^{E}_{i}(p),{\cal H}^{T}_{j}(x')]
=-i\frac{\mu}{\lambda}\delta(p)_{,i}\delta(p,p')_{,j}
\end{equation}
will also give terms with 
derivatives of $\delta$-functions,
the order of factors in supermomentum operators does not
lead 
to any factor-ordering problem. Hence, one can have
a closed algebra of the generalized constraints also in the
quantized theory, and therefore ${\cal H}_{sX}\Psi=0$
and ${\cal H}_{X}\Psi=0$ can be satisfied simultaneously [10].
The same conclusion can also be obtained in the
quantum-mechanical description of cosmodynamics. Thus, the
quantization of the extended formalism of both 
geometrodynamics and cosmodynamics leads to no problem
with a hermitian order of operators. The issue of
quantizing the gravitational 
field may then be persued
without restricting to domains where the factor-ordering
problem is circumvented or replacing the dynamical content
of Eqns. (4.19) and (4.20) for a cosmological constant.


\acknowledgements

\noindent 
This work was supported by DGICYT 
under Research Project N\mbox{$^{\underline{o}}$} PB94-0107-A.

\pagebreak

\begin{center}
{\bf Legends for figures}
\end{center}

\noindent Fig. 1: Relation of points of the two hypersurfaces
which carry the same intrinsic label and their location in
spacetime and momentum-energy sheets.

\vspace{.5cm}

\noindent Fig. 2: Related changes of the normals to the two
hypersurfaces when each of these hypersurfaces is displaced
an infinitesimal amount.

\end{document}